\documentclass{PoS}

\usepackage{amsmath}

\def\zo{\overline{z}_1}
\def\zt{\overline{z}_2}

\def\w{\omega}

\title{Threshold resummation in rapidity for colorless particle production at LHC}

\ShortTitle{Threshold Resummation}

\author{Pulak Banerjee\\
        The Institute of Mathematical Sciences, Taramani, Chennai 600113, India\\
        { Homi Bhabha National Institute, Training School Complex, Anushakti Nagar, Mumbai 400085, India}\\
        E-mail: \email{bpulak@imsc.res.in}}
        
\author{\speaker{Goutam Das}\\
        Theory Group, Deutsches Elektronen-Synchrotron (DESY),\\
         Notkestrasse 85, D-22607 Hamburg, Germany\\
        E-mail: \email{goutam.das@desy.de}}
        
\author{Prasanna K. Dhani\\
        The Institute of Mathematical Sciences, Taramani, Chennai 600113, India\\
        { Homi Bhabha National Institute, Training School Complex, Anushakti Nagar, Mumbai 400085, India}\\
        E-mail: \email{prasannakd@imsc.res.in}}
\author{V. Ravindran\\
        The Institute of Mathematical Sciences, Taramani, Chennai 600113, India\\
        { Homi Bhabha National Institute, Training School Complex, Anushakti Nagar, Mumbai 400085, India}\\
        E-mail: \email{ravindra@imsc.res.in}}
\abstract{
 We present a formalism to resum large threshold logarithms to all orders in perturbative QCD for the rapidity distribution of any colorless particle at the hadron colliders.  Using the derived resummed coefficients in two dimensional Mellin space, we present the rapidity distributions for the Higgs as well as for the Drell-Yan production to NNLO+NNLL accuracy at the LHC.
The resummed distributions give stable prediction against the variation of unphysical renormalisation and factorisation scales in both the cases. Perturbative convergence is also improved with the inclusion of the resummed result.}

\FullConference{Loops and Legs in Quantum Field Theory (LL2018)\\
		29 April 2018 - 04 May 2018\\
		St. Goar, Germany\\
		Preprint: DESY 18-115}

\begin{document}

\section{Introduction}
Resummation of large logarithms for rapidity distribution has been an interesting topic over the years and several results are already available to a very good accuracy for different processes. The fixed order (f.o) predictions are often not reliable in certain regions of phase space where large logarithms of some kinematic variables appear. For example, at the partonic threshold, where the initial partons have just enough energy to produce the final state such as a Higgs boson or $Z/W^{\pm}$ boson or a pair of leptons in addition to  soft gluons, the phase-space available for the gluons become severely constrained which results in large logarithms. In a truncated f.o calculation, these large logarithms give unreliable result and needs to be systematically resummed to all orders in perturbation theory for reliable predictions.

When talking about resummation of rapidity, 
two distinct approaches can be observed in QCD.
One we call Catani \& Trentadue approach (or Mellin-Mellin (M-M) approach)  \cite{Catani:1989ne} which was proposed for the $x_F$ distribution but can easily be extended to rapidity distribution. In this approach threshold limit is taken using both partonic scaling variable $z_1,z_2$ simultaneously going to threshold limit $1$. This basically resums all the delta ($\delta(1-z_i)$) and distributions ($\big[\frac{\ln^{n}(1-z_i)}{1-z_i}\big]_+$) arising in $z_1$ and $z_2$. Using this approach lepton pair resummation is performed  at NLL accuracy \cite{Westmark:2017uig}. The other approach, we call by Laenen \& Sterman approach (or Mellin-Fourier (M-F) approach) \cite{Laenen:1992ey}. Here partonic cross-section is written in terms of scaling variable $z$ and partonic rapidity $y_p$ and finally threshold limit is taken {\it only} for $z\to1$ which resums delta ($\delta(1-z)$) and distributions ($\big[\frac{\ln^{n}(1-z)}{1-z}\big]_+$) in $z$. However for partonic $y_p$ only delta ($\delta(y_p)$) piece is taken. Using this approach, resummation has been performed for $W^{\pm}$ production \cite{Mukherjee:2006uu} as well as Drell-Yan (DY) rapidity upto NNLL accuracy \cite{Bolzoni:2006ky, Bonvini:2010tp}.

We follow the M-M approach and derive an all order resummed result in two dimensional Mellin space for rapidity distribution of any colourless state $F$ that can be produced in hadron colliders. We  present our results in terms of Mellin variables $N_1$ and $N_2$ corresponding to $z_1$ and $z_2$ respectively. In the Mellin space, the limits $z_i \rightarrow 1$ translate into $N_i \rightarrow \infty$ and large logarithms proportional to $\ln(N_i)$ are resummed to all orders in perturbation theory. We present numerical results for resummed rapidity distributions for Higgs \cite{Banerjee:2017cfc} and DY \cite{Banerjee:2018vvb} productions at the LHC.

\section{Theoretical Framework}
The rapidity distribution of a colorless state $F$ can be written as
\begin{eqnarray}\label{sighad}
{d \sigma^I\over dy } &=&
\sigma^I_{\rm B}(x_1^0,x_2^0,q^2) 
\sum_{ab=q,\overline q,g}
\int_{x_1^0}^1 {dz_1 \over z_1}\int_{x_2^0}^1 {dz_2 \over z_2}~ 
\!\!\times
{\cal H}^I_{ab}\left({x_1^0 \over z_1},{x_2^0\over z_2}
\right)
\Delta^I_{d,ab} (z_1,z_2,q^2).
\end{eqnarray}
For brevity,  the renormalization scale ($\mu_R$) and the factorisation scale ($\mu_F$) dependences are kept implicit in the above equation. Here the hadron level rapidity  is $y={1 \over 2} \ln(p_2.q/p_1.q)={1 \over 2 } \ln\left({x_1^0/x_2^0}\right)$; $\tau=q^2/S=x_1^0 x_2^0$, $q$ being the momentum of the final state $F$, $S=(p_1+p_2)^2$,  where $p_i$ are the momenta of incoming hadrons $P_i~(i=1,2)$. For the DY process, the state $F$ is a pair of leptons with invariant mass $q^2$ ($I=q$), $\sigma^I=d\sigma^{q}(\tau,q^2,y)/dq^2$ whereas for the Higgs boson production through gluon (bottom anti-bottom) fusion, $I=g(b)$ and $\sigma^I=\sigma^{g(b)}(\tau,q^2,y)$. The luminosity  ${\cal H}^I_{ab}$ in Eq.{\ref{sighad}} is given by the product of
 parton distribution functions (PDFs)
 $f^{P_1}_a(x_1,\mu_F^2)$ and $f^{P_2}_b(x_2,\mu_F^2)$,  renormalized at $\mu_F$.  The partonic coefficient functions denoted by $\Delta^I_{d,ab}$ depend on the parton level scaling variables $z_i, (i=1,2)$. Using factorization properties of the cross sections and renormalization group invariance, the threshold enhanced contribution to the $\Delta_{d,ab}^I$ denoted by $\Delta^{\rm SV}_{d,I}$ was shown to exponentiate  \cite{Ravindran:2006bu} as  
\begin{align}\label{delta}
\Delta^{\rm SV}_{d,I} ={\cal C} \exp
\Big({\Psi^I_d(q^2,\mu_R^2,\mu_F^2,\zo,\zt,\epsilon)}\Big)\, \Big|_{\epsilon = 0} \,,
\end{align}
where the exponent $\Psi^I_d $ is both ultraviolet and infrared finite to all orders in perturbation theory. It contains finite distributions computed in $4+\epsilon$ space-time dimensions expressed in terms of two shifted scaling variables $\zo = 1- z_1$ and $\zt = 1-z_2$ and takes the following form:
\begin{align} \label{psi}
\Psi^I_d &=
\Big(
\ln \Big(Z^I(\hat a_s,\mu_R^2,\mu^2,\epsilon)\Big)^2
+ \ln \big|\hat F^I(\hat a_s,Q^2,\mu^2,\epsilon)\big|^2
\Big)
\delta(\zo) \delta(\zt)
\nonumber\\
& - {\cal C} \Big( \ln \Gamma_{II}(\hat a_s,\mu^2,\mu_F^2,\zo,\epsilon)~ \delta(\zt) + (\zo \leftrightarrow \zt)    \Big) 
+ 2~ \Phi^{I}_d(\hat a_s,q^2,\mu^2,\zo,\zt,\epsilon).
\end{align}
We have defined, $Q^2=-q^2$ and the scale $\mu$ is introduced to define the dimensionless strong coupling constant $\hat a_s=\hat g_s^2/16 \pi^2$ in dimensional regularization, which is related to renormalised $a_s$ through the renormalization constant $Z (a_s(\mu_R^2))$ $i.e.$, $\hat a_s  = (\mu/\mu_R)^\epsilon Z (\mu_R^2) S_\epsilon^{-1} a_s(\mu_R^2)$, $S_\epsilon = \exp[(\gamma_E-\ln 4 \pi) \epsilon/2]$, $\gamma_E=0.57721566\cdot\cdot\cdot$ is Euler-Mascheroni constant. 
The definition of double Mellin convolution ${\cal C}$ is given in \cite{Ravindran:2006bu}, and it is understood that the regular functions resulting from various convolutions are dropped. The overall operator renormalization constant $Z^I$  renormalizes the bare form factor $\hat F^I$; the corresponding anomalous dimension is denoted by $\gamma_I$. The diagonal mass factorization kernels $\Gamma_{II}$ remove the initial state collinear singularities.  We have factored out the form factor and the mass factorization kernels in $\Delta_{d,ab}^I$ in such a way that the remaining soft distribution function $\Phi^{~I}_d$ contains only soft gluon contributions. Both the form factor $\hat F^I$ and the soft distribution function $\Phi^{I}_d$ satisfy Sudakov type differential equations (see \cite{Ravindran:2005vv,Ravindran:2006cg}) which is straightforward to solve in powers of strong coupling constant and they can be found in \cite{Ravindran:2006bu,Ravindran:2005vv,Ravindran:2006cg,Ahmed:2014uya}. In terms of these solutions we arrive at the following expression (setting $\mu_R^2=\mu_F^2$):
\begin{eqnarray}
\Psi^I_d &&= ~\delta(\overline z_2)~ \Bigg({ 1 \over \overline z_1} \Bigg\{ \int_{\mu_F^2}^{q^2 ~\overline z_1}
{d \lambda^2 \over \lambda^2}~ A_I\left(a_s(\lambda^2)\right) 
+ D^I_d\left(a_s(q^2~\zo)\right) \Bigg\}
\Bigg)_+ 
+ {1 \over 2} \Bigg( {1 \over \overline z_1 \overline z_2 }
\Bigg\{A^I(a_s(z_{12})) 
\nonumber \\
&& 
+ {d D^{I}_d(a_s(z_{12}))\over d\ln z_{12}} \Bigg\}\Bigg)_+
+ {1 \over 2}
\delta(\overline z_1) \delta(\overline z_2)
\ln \Big(g^I_{d,0}(a_s(\mu_F^2))\Big)
+ (\overline z_1 \leftrightarrow \overline z_2)
\end{eqnarray}
Here $z_{12}=q^2 \overline z_1 \overline z_2$ and $A^I$ are the cusp anomalous dimensions which are known upto four loops \cite{Moch:2018wjh}. The finite function $D^I_d$ can be expanded order by order in strong coupling and can be found from inclusive counterpart with the use of following identity \cite{Ravindran:2006bu,Ravindran:2007sv}:
\begin{align}\label{iden}
\int_0^1 dx_1^0 \int_0^1 dx_2^0 \left(x_1^0 x_2^0\right)^{N-1}{d \sigma^I \over d y}=\int_0^1 d\tau~ \tau^{N-1} ~\sigma^I\,,
\end{align}
where the $\sigma^I$ is the inclusive cross section.  
Comparing against $D^I$ from the inclusive cross section, we obtain
\begin{eqnarray}
D_{d,1}^I &=&  D^I_1 ~;~
D_{d,2}^I = D^I_2 -\zeta_2 \beta_0 A_1^I  ~;~
D_{d,3}^I = D^I_3 + \zeta_2(-\beta_1 A_1^I -2 \beta_0 A_2^I
-2 \beta_0^2 f_1^I ) - 4 \zeta_3 \beta_0^2 A_1^I
\nonumber\\
D_{d,4}^I &=& D^I_4 + \zeta_2 (-2 \beta_1 A_2^I - \beta_2 A_1^I
-\beta_0(3 A_3^I + 5 \beta_1 f_1^I)
-6 \beta_0^2 f_2^I 
-12 \beta_0^3 \overline {\cal G}_{1}^{I,1})
-{57 \over 5} \zeta_2^2 \beta_0^3 A_1^I 
\nonumber\\
&&
-\beta_0 \zeta_3 (12 \beta_0 A_2^I +10 \beta_1 A_1^I +12 \beta_0^2 f_1^I )\,.
\end{eqnarray}
After taking the double Mellin moments \cite{Catani:2003zt} of Eq. \ref{delta} we arrive at the $N_1$-$N_2$ space cross-section:
\begin{eqnarray}
\label{eqn:expG}
\tilde \Delta_{d,I}^{(res)}(N_1,N_2) \equiv \tilde \Delta_{d,I}^{SV}(\w) &=& \int_0^1 dx_1^0\big(x_1^0\big)^{ N_1-1} \int_0^1 dx_2^0 \big(x_2^0\big)^{ N_2-1}  \Delta_{d,I}^{SV}
\equiv g^I_{d,0}(a_s) \exp\big( g^I_d (a_s,\w)\big) 
\end{eqnarray}
where $\w = a_s \beta_0 \ln(\overline N_1 \overline N_2)$ (where $\overline N_i = e^{\gamma_E} N_i, i=1,2$).
Eq. \ref{eqn:expG} is organised in such a way that $g^I_d(a_s,\w)$ contains only $N_1,N_2$ dependent terms whereas $g^I_{d,0}(a_s)$ are $N_1,N_2$ independent.
The $N_i$ independent coefficients $g^I_{d,0}(a_s)$ can be expanded in powers of $a_s$ as $\ln(g^I_{d,0}) = \sum_{i=0}^\infty a_s^i  l^{I,(i)}_{g_0}$.
The exponent $g^I_d(a_s,\w)$ takes the canonical form:
\begin{eqnarray}
g^I_d(a_s,\w)=g^I_{d,1}(\w) \ln(\overline N_1 \overline N_2) + \sum_{i=0}^\infty a_s^i g^I_{d,i+2}(\w)\,.  
\end{eqnarray}
To perform resummation at NNLO+NNLL accuracy, we need  resummed coefficients upto $g^I_{d,3}$ and the prefactors upto $l^{I,(2)}_{g_0}$ and those can be found in \cite{Banerjee:2017cfc}\footnote{The $g^I_{d,4}$ and $l^{I,(3)}$ coefficients can also be found in the first arXiv version of \cite{Banerjee:2017cfc}.}.
Exponentiation of the coefficients $g^I_{d,i}$   resums the terms $a_s \beta_0 \ln(\overline N_1 \overline N_2)$ systematically to all orders in perturbation theory. The resummed result has to be properly matched with the fixed order avoiding any double counting of the logarithms. The matched cross-section takes the following form:
\begin{align}\label{eq8}
{d\sigma^{I,(res)} \over dy } =  
{d\sigma^{I,(f.o)} \over dy } +
\, &{\sigma^{I}_B } \int_{c_{1} - i\infty}^{c_1 + i\infty} \frac{d N_{1}}{2\pi i}
 \int_{c_{2} - i\infty}^{c_2 + i\infty} \frac{d N_{2}}{2\pi i} 
e^{y(N_{2}-N_{1})}
\left(\sqrt{\tau}\right)^{c_I-N_{1}-N_{2}} 
\tilde f_{I}(N_{1}) 
 \tilde f_{I}(N_{2}) 
\nonumber\\&
\hspace{-0.1cm}\times
\bigg[ \tilde \Delta_{d,I}^{(res)}(N_1,N_2) -  \tilde \Delta_{d,I}^{(res)}(N_1,N_2)\Big|_{\rm{tr}} \bigg] \,,
\end{align}
Here $c_I = -4 (I=g)$ and $2 (I=q)$. The subscript $tr$ refers to the result obtained from Eq.(\ref{eqn:expG}) by truncating at desired accuracy in $a_s$. Note that the coefficients $g_{d,0}^I$ and $g_{d,i}^I$ are functions of cusp ($A^I_i$), collinear ($B^I_i$), soft ($f^I_i$), UV ($\gamma^g_i$) anomalous dimensions and universal soft terms ${\overline\cal G}^{I,i}_{d,j}$ and process dependent constants $G^{I,i}_j$ of virtual corrections. These constants are known to sufficient order to perform resummation to NNLL accuracy. The $N_i$ dependent terms inside the square bracket appropriately multiplied with $N_i$ dependent PDFs, namely $\tilde f_I(N_i)$ have to undergo two Mellin inversions to obtain the final result in terms of $\tau$ and $y$.  We have used minimal prescription advocated in \cite{Catani:1996yz} to perform the Mellin inversion to finally get resummed rapidity distribution.
\section{Results}
\subsection{Higgs rapidity distribution}
\begin{figure}[ht]
\centerline{
\includegraphics[width=9.5cm, height=4.8cm]{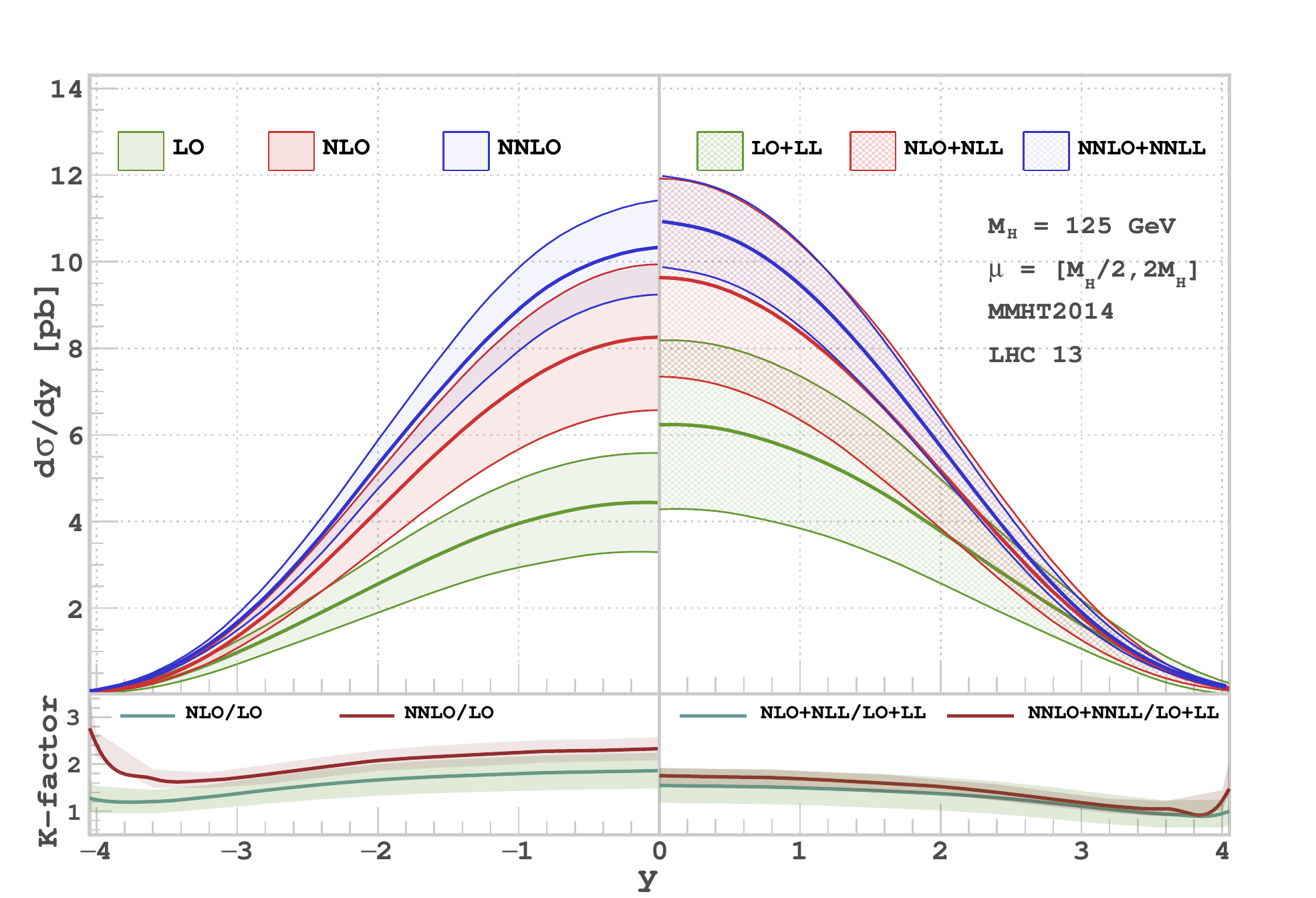}
}
\caption{Higgs rapidity distributions for fixed order (left panel) upto NNLO and resummed (right panel)  contributions upto NNLO+NNLL are presented with scale variation around central scale choice $M_H$. The respective K-factors are shown at the bottom panel.}
\label{fig1}
\end{figure}
\begin{figure}[ht]
\centerline{
\includegraphics[width=7.5cm, height=5cm]{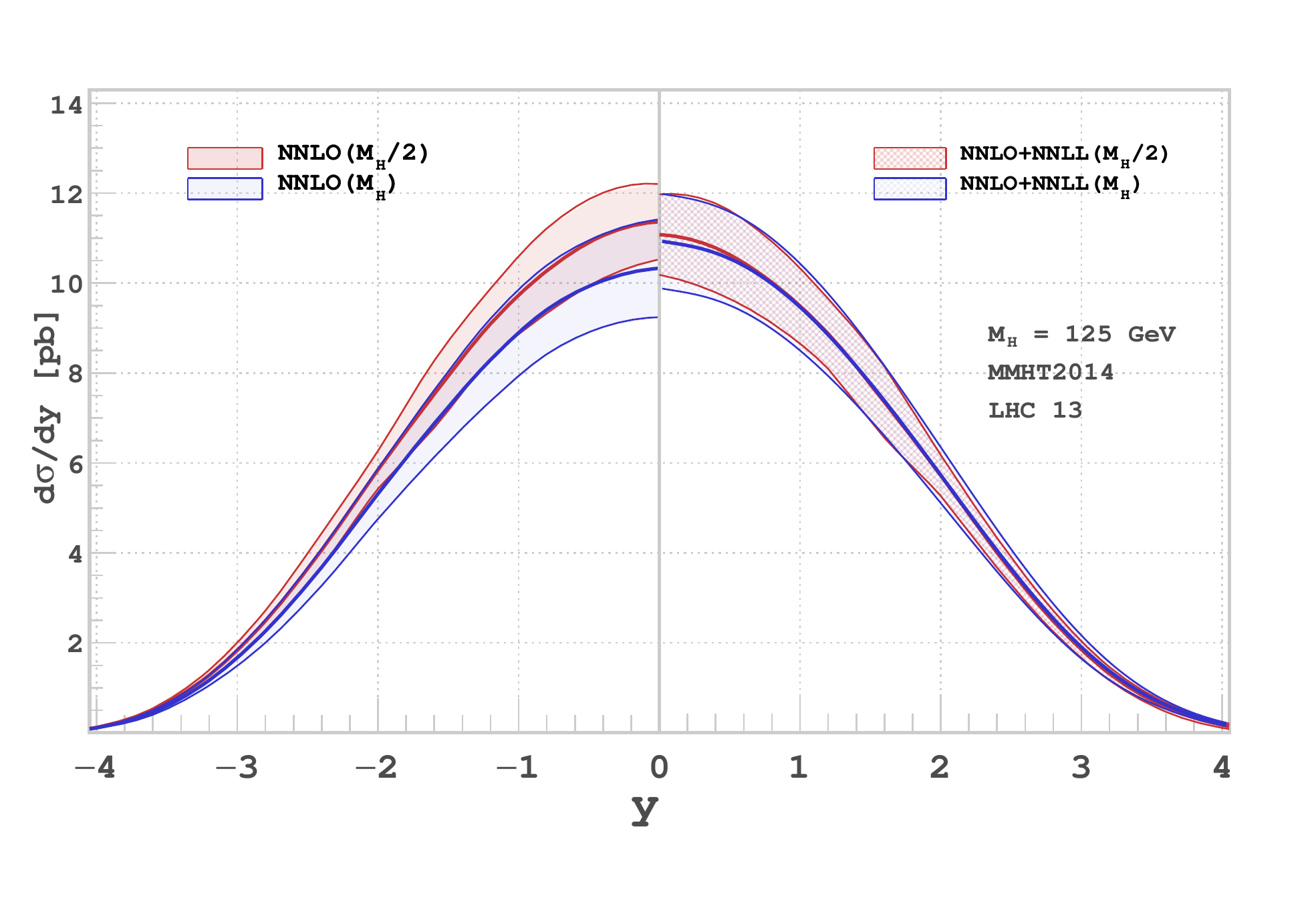}
\includegraphics[width=7.5cm, height=5cm]{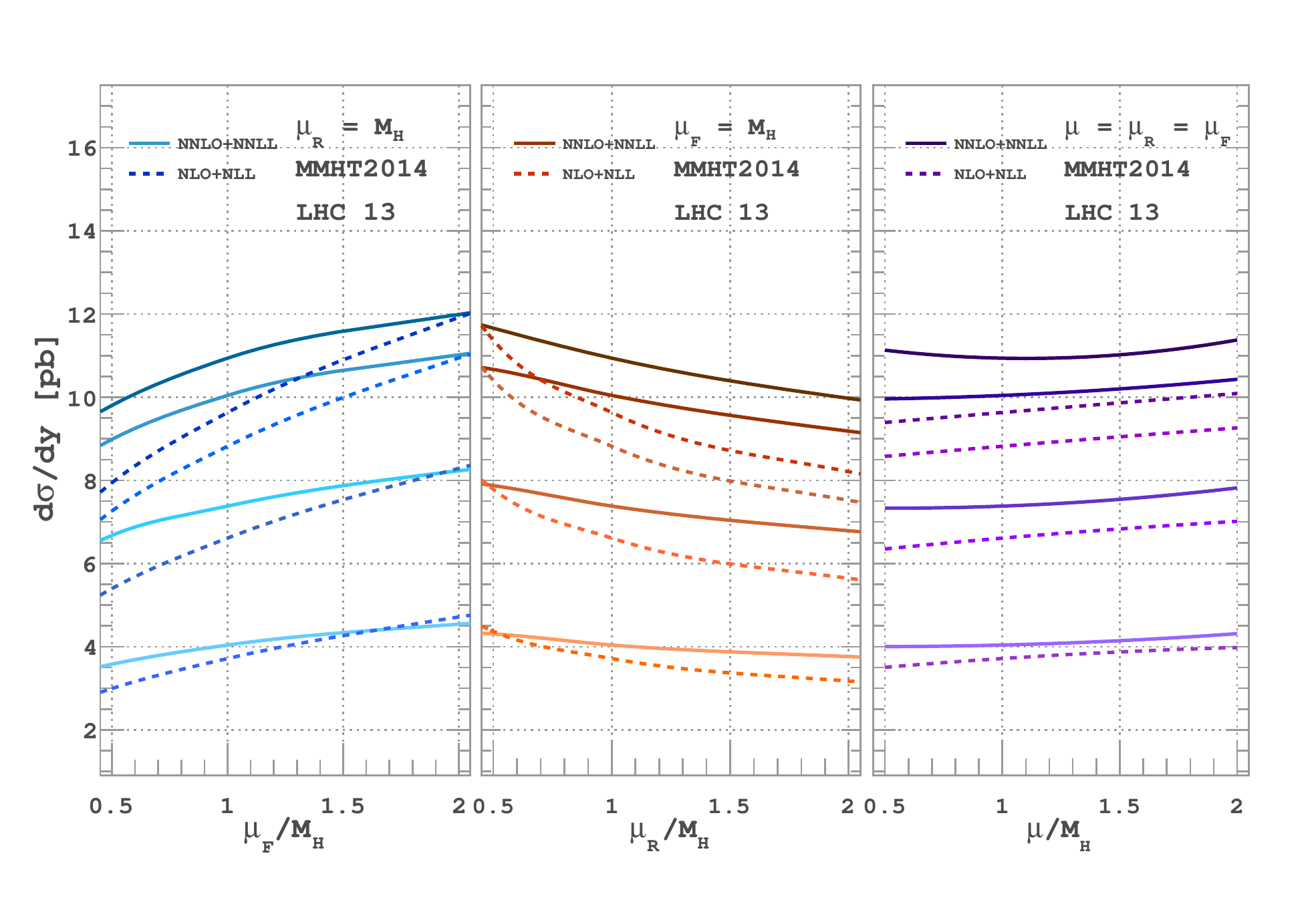}
}
\caption{
(Left) Higgs rapidity distributions for fixed order and resummed contributions
are presented with scale variation around central scale choices $M_H/2$ and $M_H$ at NNLO+NNLL.
(Right) $\mu_F$, $\mu_R$ scale variations for different benchmark $y$ values (starting from the top $y=0, 0.8, 1.6, 2.4$).}
\label{fig10}
\end{figure}

To perform numerical analysis for the Higgs rapidity distribution, we have adopted following choice of parameters: $\sqrt{S} = 13$ TeV, $M_H = 125$ GeV, $n_f = 5$, $M_t = 173$ GeV and used MMHT2014 \cite{Harland-Lang:2014zoa} PDF set with corresponding value of strong coupling constant at each order in perturbation theory. While f.o results up to NNLO are obtained using publicly available code FEHIP \cite{Anastasiou:2005qj}, the resummed contributions are included up to NNLL using an in-house Fortran code. To assess remaining scale uncertainty due to unphysical renormalisation and factorisation scale, we vary them between $[M_H/2,2M_H]$ around the central scale $\mu_R = \mu_F = M_H$ with the constraint $1/2\leq\mu_R/\mu_F\leq2$. In Fig.\ \ref{fig1}, we have plotted production cross section for the Higgs boson as a function of its rapidity $y$ up to NNLO in left panel and to NNLO+NNLL in right panel along with respective $K$-factors. We observe (see Fig.\ \ref{fig1}) that  the extent of overlap between consecutive orders in resummed case is better compared to fixed order indicating the fact that inclusion of higher order corrections has improved the convergence of the perturbation series. In particular, NNLO+NNLL increases approximately by $13\%$ with respect to NLO+NLL whereas corresponding number for NNLO over NLO is approximately $25\%$. We also found that the choice of different central scales has minimum effect on the resummed result at NNLO+NNLL level (see Fig.\ \ref{fig10} left). 
The scale uncertainties coming from the variation of $\mu_F$ and $\mu_R$ are also reduced by the inclusion of resummed contributions (Fig.\ \ref{fig10} right).
\subsection{Drell-Yan rapidity distribution}
\begin{figure}[b]
\includegraphics[width=7.5cm,height=4.85cm]{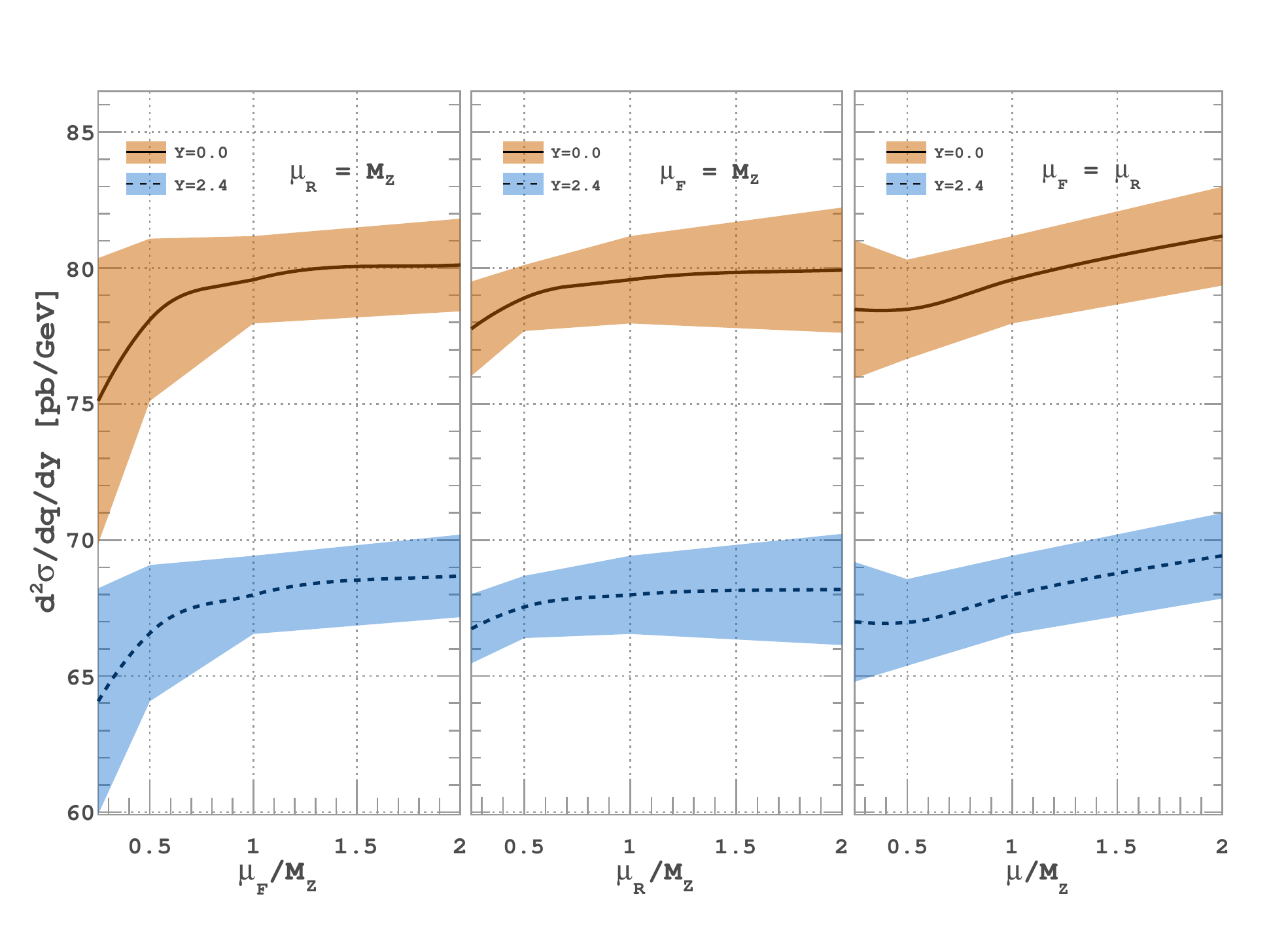}
\includegraphics[width=7.5cm,height=4.85cm]{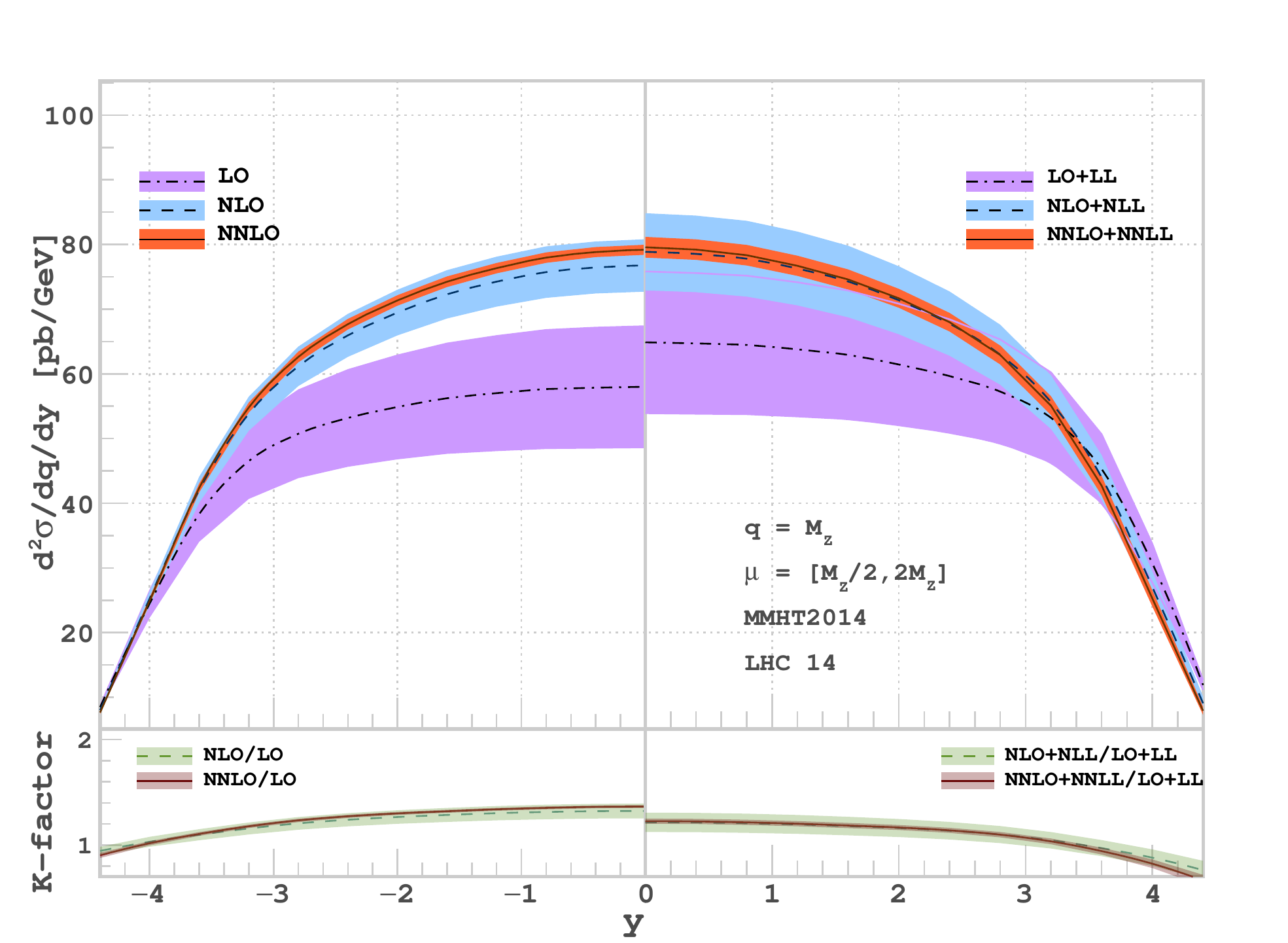}
\caption{
(Left) DY cross sections against $\mu_F$(left), $\mu_R$(middle) and $\mu$(right) variations at NNLO+NNLL 
for 14 TeV LHC.
(Right) Rapidity distribution for 14 TeV LHC at $q=M_Z$ with bottom panels representing the K-factors.}
    \label{fig2}
\end{figure}
\begin{figure}
\includegraphics[width=7.5cm, height=5.5cm]{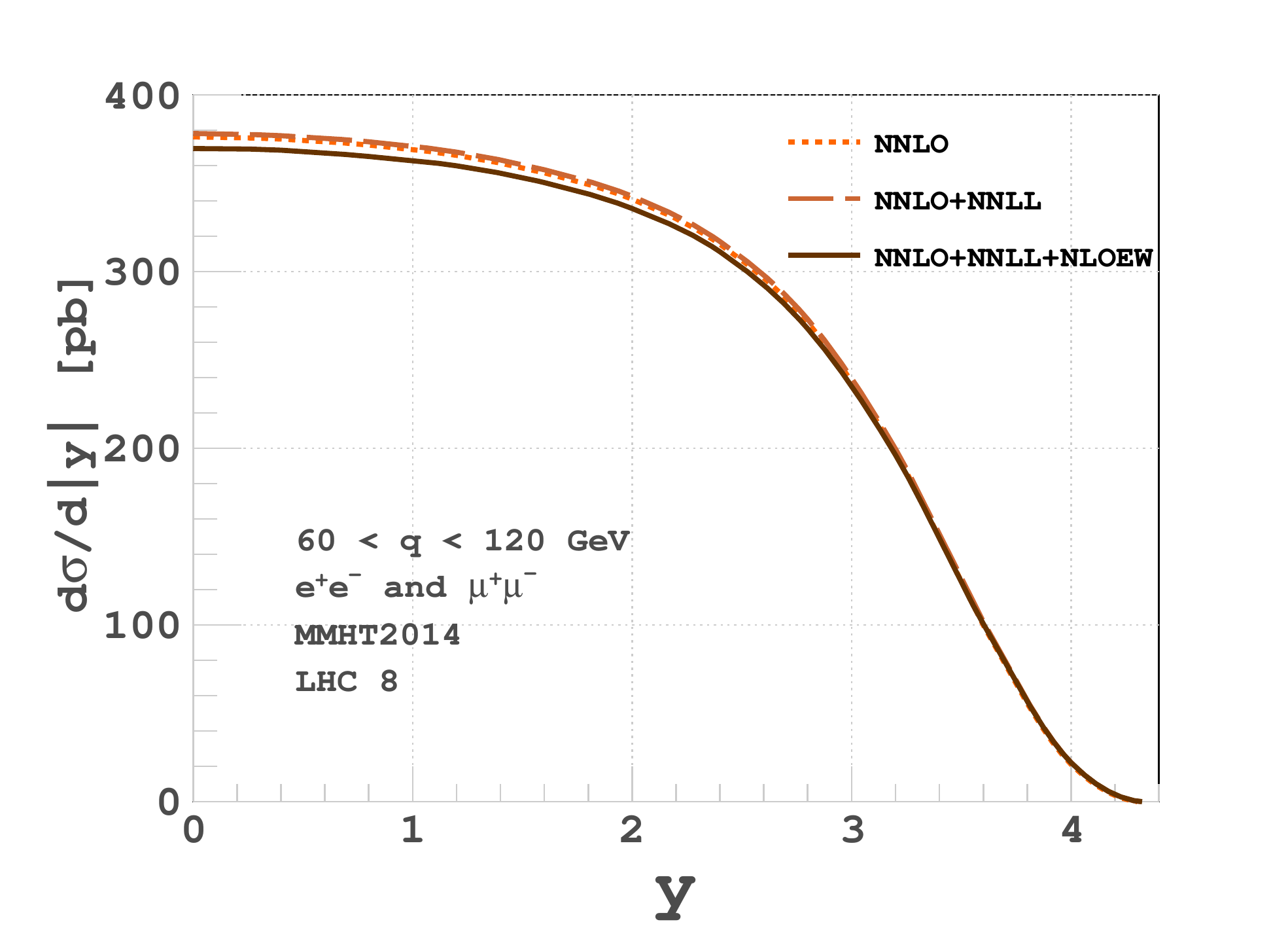}
\includegraphics[width=7.5cm, height=5.5cm]{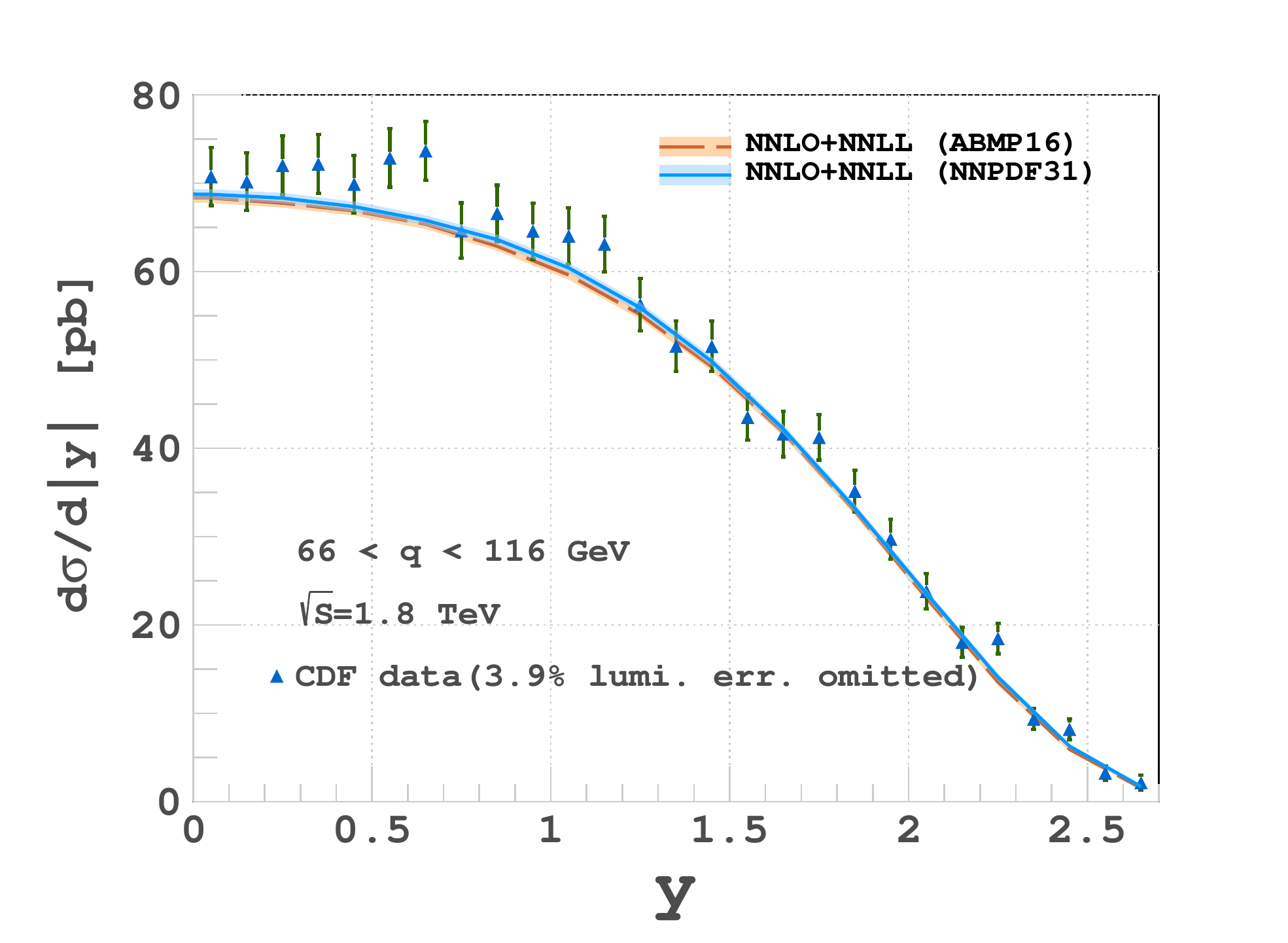}
\caption{
(Left) DY rapidity distribution at NNLO+NNLL for 8 TeV LHC in the invariant mass range $60<q<120$ GeV.
(Right) Comparison between resummed results and the CDF data~\cite{Affolder:2000rx} at $\sqrt{s}=1.8$ TeV in the invariant mass range  $66<q<116$ GeV for two different PDF sets.
}\label{fig6}
\end{figure}

For DY rapidity distribution we choose to work at 14 TeV LHC and focus mainly the $Z$-peak region. The NNLO contributions are obtained from Vrap-0.9 \cite{Anastasiou:2003ds}. 
We performed a detailed analysis on the choice of central scale and found the best prediction for the f.o case is $(\mu_r,\mu_f) = (1,1)M_Z$ whereas in resummed case it is $(\mu_r,\mu_f) = (1/2,1)M_Z$ (see Fig.\ \ref{fig2} left). In DY case also we see a better perturbative convergence compared to the f.o. The scale uncertainty however is more in the resummed case compared to the f.o (Fig.\ \ref{fig2} right). The reduced scale uncertainty at f.o is due to the large cancellation of the contributions from different partonic channels which could be accidental and might not hold at higher orders. Resummation only takes care of the large logarithms coming from the distribution in the $q\bar{q}$ channel; therefore considering only $q\bar{q}$ channel, we get  less scale uncertainty compared to the f.o as expected. The PDF uncertainties are also consistent among different groups and remains within $2\%$ at NNLO+NNLL. We also made a numerical comparison between the M-F and M-M approaches keeping parameters for both cases same as in \cite{Bonvini:2010tp}. We found a significant difference at LO+LL level; though at higher orders the differences are not much at the level of cross-section. The M-M approach however provides a better perturbative convergence ( see Table-\ref{table1}). Finally we stress that at this accuracy the electro-weak (EW) corrections are important. Using publicly available code Horace \cite{CarloniCalame:2007cd} we have included the EW corrections at NLO accuracy with $q$-integrated NNLO+NNLL QCD result at $8$ TeV LHC (Fig.\ \ref{fig6} left). Moreover we compare our prediction with CDF data \cite{Affolder:2000rx} for $\sqrt{S}=1.8$ TeV integrated over $q$ in the range $66 < q < 116$ GeV and find a very good agreement (Fig.\ \ref{fig6} right).

\begin{center}
\begin{table*}
\label{table1}
 \renewcommand{\arraystretch}{1.}
\begin{tabular}{|p{1.2cm}||p{1.cm}|p{1.cm}|p{1.cm}||p{1.cm}|p{1.cm}|p{1.1cm}||p{1.cm}|p{1.1cm}|p{1.2cm}|}
 \hline
 $(\frac{\mu_R}{M_Z}, \frac{\mu_F}{M_Z})$ & $ \text{{\footnotesize LO}}$ &$\text{{\footnotesize LL}}_{\text{{\tiny M-F}}}$ & $ \text{{\footnotesize LL}}_{\text{{\tiny M-M}}}$
    &$\text{{\footnotesize NLO}}$ & $\text{{\footnotesize NLL}}_{\text{{\tiny M-F}}}$ & $\text{{\footnotesize NLL}}_{\text{{\tiny M-M}}}$ & $\text{{\footnotesize NNLO}}$ & $\text{{\footnotesize NNLL}}_{\text{{\tiny M-F}}}$  & ${ \text{{\footnotesize NNLL}}_{\text{{\tiny M-M}}}}$ \\
\hline
\hline
 (2,\,2) & 72.626 &+0.988  & +3.219  & 73.450 &+1.639& +1.796 & 70.894 &+0.630 & +0.646  \\
 \hline
 (2,\,1) & 63.197 &+0.768 & +2.595 & 70.625 &+0.761   &+1.017& 70.360 &+0.292 & +0.317\\
  \hline
 (1,\,2) &  72.626 &+1.095 & +3.577 & 73.535 &+1.912 & +1.760  & 70.509 &+0.510 & +0.395 \\
  \hline
 (1,\,1)  & 63.197 &+0.851 &+2.887& 71.395 &+0.858 & +0.901 & 70.537 &+0.248 & +0.167 \\
 \hline
 (1,\,0.5) & 53.241 &+0.621 & +2.216 & 67.581 &+0.156&+0.140 & 69.834 &- 0.001 & - 0.094  \\
  \hline
 (0.5,\,1)& 63.197 &+0.953 &+3.278& 72.355 &+0.945 &+0.681 & 70.266 &+0.091& - 0.015 \\
 \hline
 (0.5,\,0.5) & 53.241 &+0.695 &+2.504& 69.259 &+0.102 &  - 0.154 & 70.283 &- 0.039& - 0.146 \\
 \hline
\end{tabular}
\caption{ Comparison of resummed results between M-F and M-M approach in the minimal prescription scheme at $y=0$ for various choices of scales.} \label{table1}
\end{table*}
\end{center}

%
\section{Conclusion}
We have developed a formalism to resum threshold logarithms in double Mellin space for the rapidity distribution of a colorless final state $F$ produced at the hadron collider. An analytic expression of the resummed coefficients upto N$^3$LL has been presented in terms of double Mellin variables $N_1$ and $N_2$. As an application we have studied the role of the resummed threshold logarithms for the rapidity distribution for Higgs and DY productions at the LHC. We have performed a detailed study on scale variations and central scale choice as well as estimated uncertainty coming from PDFs. Numerical impact of our resummation in double Mellin space has significant differences at the leading logarithmic accuracy compared to the existing results in literature; however we found agreement at NNLO+NNLL level. Our resummed coefficients can be used for rapidity distribution of any colorless final state produced at the LHC.  The numerical analysis presented here would be useful to understand the properties of the Higgs boson as well as will be very useful for precise determination of PDFs at the LHC.

\bibliography{ll2018GD}
\bibliographystyle{JHEP}
\end{document}